\documentclass[prl,twocolumn]{revtex4-1}
\usepackage{epsfig}

\def\beq{\begin{equation}}
\def\eeq{\end{equation}}
\def\be{\begin{equation}}
\def\ee{\end{equation}}
\def\bea{\begin{eqnarray}}
\def\eea{\end{eqnarray}}

\def\la{\langle}
\def\ra{\rangle}
\def\ms{\overline{MS}}
\def\t{\tilde }
\def\Lam{\Lambda}
\newcommand{\qq}{\langle \bar q q \rangle} 

\newcommand{\gsim}{\raisebox{-0.13cm}{~\shortstack{$>$ \\[-0.07cm] $\sim$}}~}

\begin{document}

\title{$\Lam^{\rm QCD}_{\ms}$ 
from Renormalization Group
Optimized Perturbation}

\author{Jean-Lo\"{\i}c Kneur}
\author{Andr\'e Neveu}
\affiliation{CNRS, Laboratoire Charles Coulomb UMR 5221, F-34095, Montpellier, France}
\affiliation{Universit\'e Montpellier 2, Laboratoire Charles Coulomb UMR 5221, F-34095, Montpellier, France}

\begin{abstract}
A recent extension of a variationally optimized perturbation, 
combined with renormalization group properties in a straightforward way,  
can provide approximations to nonperturbative quantities  
such as the chiral symmetry breaking order parameters typically. 
We apply this to evaluate, up to third order in this modified perturbation, the ratio
$F_\pi/\Lambda_{\ms}$, where $F_\pi$ is the pion decay constant and $\Lambda_{\ms}$ the basic
QCD scale in the $\ms$ scheme. Using experimental $F_\pi$ input value we obtain 
$\Lam^{n_f=2}_{\ms} \simeq 255_{-15}^{+40} $ MeV, where quoted errors are estimates of theoretical
uncertainties of the method. This compares reasonably well with some recent lattice simulation results. 
We briefly discuss prospects (and obstacles) for extrapolation to $\alpha_S(\mu)$  at perturbative $\mu$ values.
\end{abstract}
\maketitle
In the chiral symmetric, massless quarks limit, the strong coupling 
$\alpha_S(\mu)$ at some reference scale $\mu$
is the only QCD parameter. Equivalently the Renormalization-Group (RG) invariant scale
\be
\Lam^{n_f}_{\ms}\equiv \mu e^{-\frac{1}{\beta_0\alpha_S}} (\beta_0\alpha_S)^{-\frac{\beta_1}{2\beta^2_0}}\:(\cdots)\;,
\label{Lamdef1}
\ee
in a specified renormalization scheme, is the fundamental QCD scale. In (\ref{Lamdef1}) $\beta_0$, $\beta_1$ are
(scheme-independent) one- and two-loop RG beta function coefficients, and ellipsis denote 
higher orders scheme-dependent RG corrections as will be specified below.
As indicated $\Lambda^{n_f}_{\ms}$ also depends on the number of active quark flavors $n_f$,  
with non-trivial (perturbative) matching relations at the quark mass thresholds (see
e.g. the QCD chapter in (\cite{PDG}) for a recent review and original references).  
$\alpha_S$ has been extracted from many different 
observables confronted with theoretical predictions, and its present World average is 
impressively accurate~\cite{PDG}: $\alpha_S(m_Z)\simeq .118\pm .001$ 
(though there are long-standing tensions with 
values from structure functions~\cite{alphaS_DIS}: $\alpha_S(m_Z)\simeq .114\pm.002$).  
In any case, it is still of great interest to estimate $\Lam_{\ms}$
from other observables, and other theoretical methods, specially to access
the deep infrared, nonperturbative QCD regime for $n_f=2 (3)$, where a perturbative extrapolation
from $\alpha_S(m_Z)$ is unreliable.
Indeed for several years determinations 
of $\Lam^{n_f}_{\ms}$ for $n_f \le 2(3)$ from Lattice simulations  
have been the subject of much activities.     

In this letter we explore a different route to 
estimate such quantities, more rooted in perturbation theory, and where  
the dynamically broken chiral symmetry due to the light $u$, $d$ (and $s$) quarks plays a crucial role. 
The main order parameters of chiral symmetry breaking, namely the chiral quark condensate $\qq$ and pion decay
constant $F_\pi$, should be entirely determined by the unique scale $\Lam_{\ms}$ in the strict chiral limit.   
Well-established arguments usually consider hopeless 
to calculate the above order parameters from
QCD first principle, except on the lattice. First, most obviously because of the mentioned 
nonperturbative regime at the relevant scale close to $\Lam_{\ms}$,   
implying a priori large $\alpha_S$ values 
invalidating reliable perturbative expansions.
Second, standard perturbative series of those quantities at arbitrary orders 
are anyhow proportional to the quark masses $m_q$ (up to powers of $\ln m_q$), 
so trivially vanish  in the strict chiral limit $m_q\to 0$. Moreover, general arguments, 
related to the problem of resumming presumed factorially divergent 
perturbative series at large orders\cite{renormalons}, 
seem to further invalidate any perturbative approach to calculate the order parameters. 
We will see how to circumvent at least the first two arguments above, by a peculiar 
modification of the ordinary perturbative expansion in $\alpha_S$, with possible 
improvements of the convergence and resummation properties of such a modified series.
In this letter we concentrate on determining $F_\pi/\Lam_{\ms}$ at successive orders of this modified
perturbation, thus extracting $\Lam_{\ms}$ values from the pion decay constant value
$F_\pi$.\\
Our method has been recently applied\cite{rgopt1} to
the $D=2$ Gross Neveu (GN) $O(2N)$ model\cite{GN}, 
which shares many properties with $D=4$ QCD: it is 
asymptotically free, has a (discrete) chiral symmetry for $m= 0$,  
dynamically broken with a fermion mass gap. The exact mass gap is known for arbitrary $N$,   
from Thermodynamic Bethe Ansatz\cite{TBAGN},   
allowing accurate tests of our method. Using only the two-loop ordinary
perturbative information, we obtained approximations to the exact mass gap at the percent
or less level~\cite{rgopt1}, for any $N$ values. 

The basic framework\cite{delta} is to 
introduce an unphysical parameter $0<\delta<1$, interpolating between ${\cal L}_{free}$ and 
${\cal L}_{interaction}$, such that the relevant fermion (quark) mass 
 $m_q$ becomes an arbitrary ``variational'' parameter: 
\beq
{\cal L}_{QCD}(m_q,\alpha_S) \to  {\cal L}_{QCD}(m (1-\delta)^a, \alpha_S \delta)
\label{LQCDdel}
\eeq
where ${\cal L}_{QCD} (\alpha_S)$ stands for  
the standard complete QCD Lagrangian,  and  $m_q$ originally is a current quark mass 
relevant for chiral symmetry breaking. 
In the following we shall mainly 
consider two quark flavors $u,d$ and the corresponding  $SU(2)_L\times SU(2)_R\to SU(2)_V$ 
chiral symmetry breakdown, with $m\equiv (m_u+m_d)/2$ neglecting as usual the $m_u- m_d$ difference.
The extra parameter $a$ in (\ref{LQCDdel}) reflects the large freedom 
in the modified interpolating Lagrangian, and will allow imposing further physical 
(or technical) constraints,  as we shall specify later. \\ 
The procedure is fully consistent with renormalizability and gauge invariance, provided
that the above redefinition of the QCD coupling $\alpha_S\to \delta \alpha_S$ is performed
consistently for all interaction terms
appropriate for gauge invariance and renormalizability. 
Working with the above Lagrangian is perturbatively equivalent
to taking any standard renormalized series in $g\equiv 4\pi\alpha_S$ 
for a physical quantity, re-expanded  
in powers of $\delta$ after substitution:
\beq m \to  m\:(1- \delta)^a,\;\; g \to  \delta \:g\;.
\label{subst1}
\eeq
One takes afterwards the $\delta \to 1$ limit to recover the original {\em massless} theory.
This expansion gives, however, a remnant $m$-dependence at any finite   
$\delta^k$-order, and $m$ can be fixed conveniently 
by an optimization (OPT) prescription\cite{pms}.
The convergence of such a procedure, which may be viewed as a particular case of 
``order-dependent mapping''\cite{odm},  
has been proven\cite{deltaconv} for the $D=1$ $\lambda \phi^4$ oscillator model. 
In renormalizable $D > 1$ models, the situation is more involved and it is difficult  
to make statements on the possible convergence properties (see however \cite{Bconv} for a particular case). 
But at least the method allows to obtain
approximations to nonperturbative quantities beyond the mean field approximation in various models,
which (empirically) appear to converge rather quickly at the first few perturbative orders.  
 
Previous attempts to use this approach in QCD gave rough estimates 
of the order parameters (dynamical ``mass gap'', $F_\pi$, $\la \bar q q\ra$)\cite{qcd1}. 
But it involved a cumbersome manner of incorporating renormalization group (RG) properties within such modified
perturbative series, difficult to generalize beyond the first or second RG order and to
other physical quantities defined by their perturbative series.  
Our new proposal introduces in contrast a much simpler marriage of OPT and RG properties\cite{rgopt1}:    
consider an ordinary perturbative expansion for a physical quantity $P(m,g)$,
after applying (\ref{subst1}) and expanding in $\delta$ 
at order $k$. In addition to the OPT equation:
\beq
\frac{\partial}{\partial\,m} P^{(k)}(m,g,\delta=1)\vert_{m\equiv \tilde m} \equiv 0\;,
\label{OPT}
\eeq  
we require the ($\delta$-modified) series to satisfies a 
standard RG equation:
\beq
  \mu\frac{d}{d\,\mu} \left(P^{(k)}(m,g,\delta=1)\right) =0 
\eeq 
where the usual RG operator
\beq
 \mu\frac{d}{d\,\mu} =
\mu\frac{\partial}{\partial\mu}+\beta(g)\frac{\partial}{\partial g}-\gamma_m(g)\,m
 \frac{\partial}{\partial m} \; 
\label{RG}
\eeq
gives zero to ${\cal O}(g^{k+1})$ when applied to RG-invariant quantities. 
(NB our normalization is $\beta(g)\equiv dg/d\ln\mu = -2b_0 g^2 -2b_1 g^3 +\cdots$ and 
$\gamma_m(g) = \gamma_0 g +\gamma_1 g^2 +\cdots$. The $b_i$ and $\gamma_i$ known up 
to 4-loop are given in~\cite{bgam4loop}).
Note that, combined with Eq.~(\ref{OPT}), the RG equation takes a reduced form:
\beq
\left[\mu\frac{\partial}{\partial\mu}+\beta(g)\frac{\partial}{\partial g}\right]P^{(k)}(m,g,\delta=1)=0
\label{RGred}
\eeq
and Eqs.~(\ref{RGred}), (\ref{OPT}) completely fix [for given $a$ values in Eq.~(\ref{subst1})] 
optimized values $m\equiv \t m$ 
and $g\equiv \t g$. 

We shall now illustrate the method concretely on a well-defined perturbative series
relevant for the pion decay constant $F_\pi$. 
One very convenient definition of $F_\pi$ 
is via the axial current correlator, known at present up to 4-loop
orders~\cite{Fpi_3loop,Fpi_4loop}. More precisely:
\beq
i \langle 0| T A^i_\mu (p) A^j_\nu (0)|0 \rangle \equiv \delta^{ij}
 g_{\mu\nu} F^2_\pi +{\cal O}(p_\mu p_\nu)\label{Fpidef}
\eeq
where the axial current is $A^i_\mu \equiv \bar q \gamma_\mu \gamma_5 \frac{\tau_i}{2} \:q$,
and in this normalization $F_\pi\sim 92.3$ MeV~\cite{PDG}.\\ 
Our starting point is thus the perturbative expansion of (\ref{Fpidef}) in the $\ms$ scheme:
\bea 
& F^2_\pi(pert) = 3 \frac{m^2}{2\pi^2} \left[ \mbox{div}(\epsilon,\alpha_S) -L +\frac{\alpha_S}{4\pi}(8 L^2
+\frac{4}{3} L +\frac{1}{6}) \right. \nonumber \\ 
&\left. +(\frac{\alpha_S}{4\pi})^2 [f_{30} L^3+f_{31} L^2 +f_{32}L +f_{33}]+{\cal O}(\alpha^3_S)
\right]  \label{Fpipert}
\eea    
where $L\equiv \ln \frac{m}{\mu}$, 
$f_{30}= \frac{304}{3} -\frac{32}{9}n_f$, 
$f_{31} = -\frac{136}{3} +\frac{32}{9}n_f$, 
and $f_{32}$ and the non-RG coefficient $f_{33}$ have more lengthy expressions 
easily extracted from related calculations in \cite{Fpi_3loop}
valid for arbitrary numbers of quark flavors.  
Recently even the ${\cal O}(\alpha^3_S)$ coefficients $f_{4i}, i=0,\cdots,4$ were obtained~\cite{Fpi_4loop}, 
that we also use in our analysis~\cite{PrivComm}. 

There is however one subtlety at this stage: 
as is well-known, at this level the calculation {\em e.g.} in dimensional regularization
of (\ref{Fpidef}) actually still contains divergent terms after mass and coupling renormalization in 
$\ms$ scheme, formally indicated as $\mbox{div}(\epsilon,\alpha_S)$ in Eq.~(\ref{Fpipert}). This simply reflects
the extra {\em additive} renormalization needed for such a composite operator. But to obtain a RG-invariant 
finite expression from (\ref{Fpipert}) the subtraction of those divergences should be performed 
consistently with RG properties. 
Now to fix this subtraction at order $k$ needs knowledge of the coefficient of the $L$ term 
(equivalently the coefficient of $1/\epsilon$ in dimensional regularization) at order
 $k+1$. 
We define the needed subtraction as a perturbative series:
\be
{\rm sub}(g,m) = m^2 \sum_{i\ge 0} s_i  g^{i-1}
\label{sub}
\ee
 with coefficients determined order by order by
\be
\mu \frac{d}{d\mu}[{\rm sub}(g,m)]\equiv {\rm Remnant}(g,m)
\ee 
where the remnant part is obtained by applying the RG operator Eq. (\ref{RG}) to the 
finite part of (\ref{Fpipert}), as the latter is not separately RG-invariant. 
Thus the (finite) quantity $F^2_\pi({\rm pert})({\rm finite})-{\rm sub}(g,m)$ is RG-invariant
at a given order. Note that (\ref{sub}) does not contain any $L$ terms and necessarily 
starts with a $s_0/g$ term to be consistent with RG invariance properties. 
(This reflects the fact that the one-loop contribution in Eq.~(\ref{Fpipert})
is of order $g^0$). Completely equivalent results are obtained~\cite{qcd1} more formally 
by working with bare expressions and establishing the required RG properties.
We obtain for instance 
$s_0 = \frac{3}{4\pi^2(b_0-\gamma_0)}$, $s_1= \frac{237 + 17n_f}{16\pi^2 (-9 + 2n_f)}$, 
and higher order $s_i$ have more lengthy expressions not given here. 

We thus  apply to (\ref{Fpipert})-(\ref{sub}) the procedure (\ref{subst1}) 
and expand at order $\delta^k$, then solving OPT and RG Eqs.(\ref{OPT}), 
(\ref{RGred}). Before coming to numerical results, some important  
remarks are in order. First, Eqs.(\ref{OPT}), (\ref{RGred}) being polynomial in $(L,g)$,
one serious drawback is that at increasing $\delta$-orders there are  
(too) many solutions, most being complex (complex conjugate in fact since all coefficients of (\ref{OPT}), 
(\ref{RGred}) are real). Now an important selection comes
about if one imposes an additional constraint that the solutions
should obey asymptotically the standard perturbative
RG behavior for $g\to 0$: 
\be
\t g (\mu \gg \t m) \sim (2b_0 \ln \frac{\mu}{\t m})^{-1}\;.
\label{rgasympt}
\ee
This is a very natural requirement, otherwise optimal solutions do not match standard perturbative
behavior. An important related remark is that, after OPT, the optimized  
mass $\t m$ is consistently ${\cal O}(\Lam_{\ms})$ (rather than $m\sim 0$). Thus 
 $\t m$ plays the role of a mass gap, such that the OPT-modified expansion $F_\pi\propto \t m$ 
is a perturbation around a 'Born-level' value of ${\cal O}(\Lam_{\ms})$ in contrast with the  
original standard perturbative expansion, $F_\pi(m\to 0)\to 0$.
In fact, at $\delta^k$-order, Eq.~(\ref{RGred}) is a polynomial of order $k+1$ in $L$, thus
exactly solvable up to third order, with full analytical control of the different solutions.
However, unlike the GN mass gap
case\cite{rgopt1}, to have at least one of the RG OPT $F_\pi$ solutions behaving as (\ref{rgasympt}) 
at any $\delta^k$-orders, requires a critical value of the parameter $a$ in the interpolation (\ref{subst1}), 
namely $a=\gamma_0/(2b_0)$. This connection with RG anomalous dimensions is not too surprising: 
similarly in other theories, e.g. $\Phi^4$ in $D=3$,
specific $a$ values occur, consistent with RG critical exponent properties,
as emphasized in \cite{kleinert}, at the same time matching real optimal solutions~\cite{bec2}.\\
We thus fix $a=\gamma_0/(2b_0)$ to determine solutions at successive $\delta$-orders.
Most solutions exhibit very odd dependence
in $g$ incompatible with (\ref{rgasympt}), as also observed very similarly in the GN model 
case~\cite{rgopt1}. This perturbative RG criteria
appears to give unique solutions (at least up to third order here considered), given in Table~\ref{tabres}.\\ 
\begin{table}[h!]
\caption{Combined OPT+RG results at successive $\delta$-order} 
\begin{tabular}{|c||c|c|c|c|c||c|}
\hline
$\delta$-order $k$ & $\frac{F^{(k)}_\pi(\t m,\t g)}{\Lam^{\rm \small PA}_{\ms}} $ &  $\t L$ & $\t \alpha_S$  \\
\hline\hline
1    &   $ 0.372 \pm 0.16 i$   &   $-0.45 \pm 0.11 i$   &  $1.01 \pm 0.08 i$ \\
\hline
2  &    $ 0.353 \pm 0.03 i$     & $ -0.52 \mp 0.69 i$     &  $0.73\pm 0.02 i$   \\
\hline
3 ($s_4,f_{44}=0$) &    $ 0.351 \pm 0.08i $    &  $-0.13\mp 0.04 i$     & $0.61 \pm 0.33 i$        \\
\hline
3 ($s_4={\rm PA}[1,2]$)&    $ 0.341 \pm 0.07i $    &  $-0.23 \mp 0.04 i$     & $0.59 \pm 0.31 i$        \\
\hline
\end{tabular}
\label{tabres}
\end{table}
But those unique well-behaving RG solutions remain complex (conjugates) for $F_\pi$. 
It can always be that other models, other physical quantities~\cite{rgopt1,bec2} 
(or very different $n_f$ values) give real solutions, but 
not for $F_\pi$ at $k\le 3$ orders here explored. Since this is unphysical, we   
can only expect acceptable solutions of behavior (\ref{rgasympt}) to have at least ${\rm Re}(\t g)>0$ 
and ${\rm Im}(F_\pi) \ll {\rm Re}(F_\pi)$, the imaginary part 
indicating an intrinsic theoretical uncertainty of the method, as will be specified below. \\ 
To compare our results with other (principally Lattice) calculations, one should be
careful to use the same conventions for $\Lam_{\ms}$. We mainly use a convenient 
(Pad\'e Approximant) 3-loop form, cf.~\cite{LamlattWilson}:
 \be 
\Lam^{\rm \small PA}_{\ms} \equiv 
\mu \,e^{-\frac{1}{2b_0\,g}}\:\left(\frac{b_0\:g}{1+(\frac{b_1}
{b_0}-\frac{b_2}{b_1})\;g}\right)^{-\frac{b_1}{2b^2_0}}\;.
\ee
We also compare with a more standard 4-loop perturbative form~\cite{PDG}, with $b_3\ne 0$, 
which gives a systematic $\sim 2$ \% lower $\Lam_{\ms}$ values for our optimal $\t \alpha_S$ values. 

Comparing second and first $\delta$-orders in Table~\ref{tabres}, one observes that
the solution has a much smaller imaginary part, and also 
${\rm Re}\,\t\alpha_S$ decreases to reasonably perturbative values as the $\delta$-order increases. 
At third order, the $g^3 s_4$ term in (\ref{sub}) needs knowledge of the presently unknown 5-loop
coefficient of $L$. We have thus estimated $s_4$ either with a Pad\'e Approximant
${\rm PA}[1,2]$ from lower orders, or alternatively simply ignoring $s_4, f_{44}=0$, retaining only 
4-loop RG $\ln^p(m/\mu)$ coefficients. 
The difference between those two choices in Table~\ref{tabres} 
gives one estimate of higher order uncertainties.
We also incorporate additional theoretical uncertainties
by solving Eq.~(\ref{RGred}) truncated to lower $g$ orders, or neglecting $b_3$, etc (since RG-invariance is only 
required up to ${\cal O}(g^{k+1})$ terms at order $k$). 
Optimal RG solutions are remarkably stable with respect to such approximations on 4-loop order and 
RG truncations, with at most $\sim$ 2-3\% differences on $\Lam^{n_f=2}_{\ms}$.
In addition, as above mentioned we take into account a more intrinsic error: 
given the (unphysical) imaginary parts of the solutions,  
we empirically take the range spanned by 
 ${\rm Re}(F_\pi(\t g,\t L))- F_\pi({\rm Re}(\t g),{\rm Re}(\t L))$, as this tends to maximize the uncertainty 
for increasing ${\rm Im}(\t g,\t L)$. This gives only about 
a 1-2\% variation on $\Lam_{\ms}$ at ${\cal O}(\delta^2)$; but a larger $\sim 10$\% one at 
${\cal O}(\delta^3)$ due to the larger imaginary part of the solutions, 
perhaps an artefact of the unknown exact $s_4$ coefficient at this order. 
Since optimal solutions
are complex conjugates, another estimate could be simply 
to compare their real parts with their modulus, which gives a much more moderate difference 
(2\% at ${\cal O}(\delta^3)$). Clearly the occurrence of complex solutions is our main source 
of theoretical uncertainties, so there is potentially
room for improvements, {\it e.g.} from other more general prescriptions~\cite{bec2}. However, we
prefer to keep a conservative estimate of theoretical errors at this stage.  

Finally we can subtract out the explicit chiral symmetry breaking effects from small 
$m_u, m_d \ne 0$: it is in principle 
possible to incorporate those effects within the variational framework~\cite{qcd1}, 
but for the time being we shall simply rely on other known results. Defining $F$ as usual as the $F_\pi$
value in the strict chiral limit $m_u, m_d \to 0$, Lattice simulations recently obtained~\cite{LattFLAG}:
 $\frac{F_\pi}{F} \sim 1.073 \pm 0.015$, that we accordingly take into account in the final $\Lam^{n_f=2}_{\ms}$
numerical value. 
With all theoretical uncertainties (linearly) combined we obtain: 
\be
\Lam^{n_f=2}_{\ms} \simeq 255\pm 15^{+25}\;{\rm  MeV}\;.
\ee
The central value corresponds to ${\rm Re}\,F^2_\pi(\t g,\t L)$, the first errors
encompass both higher order and $F_\pi/F$ above mentioned uncertainties,  
while the upper bound corresponds
to $F^2_\pi({\rm Re}(\t g),{\rm Re}(\t L))$. \\
One may compare this with three main classes of lattice calculations based on very different methods.
First in the  Schr\"odinger functional scheme\cite{LamlattSchroed}, 
 $\Lambda^{n_f=2}_{\ms} = 245\pm 16(\mbox{stat})\pm16(\mbox{syst})$ MeV. Next for 
Wilson fermions \cite{LamlattWilson}:
 $\Lambda^{n_f=2}_{\ms} = 261 \pm 17 \pm 26$ MeV. Finally
for twisted fermions (including nonperturbative power corrections in analysis) \cite{Lamlatttwisted}:
$\Lambda^{n_f=2}_{\ms} = 330 \pm 23 \pm 22 _{-33}$ MeV.
Those differences are presumably related to different dynamical quark mass values in different Lattice
calculations, and also different chiral extrapolation methods (see {\it e.g.} the discussion in~\cite{Lamlatttwisted}).

Finally we could in principle extrapolate to $\alpha_S(\mu)$ at high (perturbative) scale $\mu$.
A main obstacle however is going from $n_f=2 $ to $n_f=3$, crossing the strange quark $m_s$  threshold
in the deep infrared regime, where one cannot use standard perturbative extrapolation.
But (\ref{Fpipert}) being known for arbitrary $n_f$, we can calculate similarly 
$F_\pi/\Lambda^{n_f=3}_{\ms}$. The details are skipped for
elsewhere, but the outcome is a mild variation, with $F_\pi/\Lam^{n_f=3}_{\ms} \gsim F_\pi/\Lam^{n_f=2}_{\ms}$ 
by only a few percent. 
However, the final value of $\Lam^{n_f=3}_{\ms}$ is much dependent also on the ratio $F_\pi/F_0$, where $F_0\equiv 
F_\pi(m_u,m_d,m_s\to 0)$. The amount of explicit chiral symmetry breaking from $m_s\ne 0$ 
is clearly more important than in the $n_f=2$ case, and 
indeed still subject to intense debates, with still large uncertainties even from 
Lattice results~\cite{LattFLAG}. 
Moreover, since our RG-improved OPT modifies perturbative expansions, 
it should also be used consistently to extrapolate to higher scales, which can differ substantially
from a standard perturbative extrapolation. For both those reasons we refrain from giving a 
precise prediction of $\alpha_S(m_Z)$ at this stage, mainly due to the large uncertainties 
involved in subtracting out explicit chiral symmetry breaking from $m_s$. We plan in the next future 
to implement those effects directly within the OPT framework.   

In conclusion, a straightforward implementation of RG properties
within a variationally optimized perturbation was proposed, using only perturbative information 
at the first few orders. 
In QCD, calculations at first $\delta$-order give already very reasonable approximations
to $F_\pi/\Lambda_{\ms}$, and second and (approximate) third order results
exhibit a remarkable stability. 
These results compare reasonably well with recent lattice calculations of $\Lam_{\ms}$,
though the best with those in~\cite{LamlattWilson}. We conjecture that the
remaining discrepancies (with other lattice results, and possibly with the World average $\alpha_S(m_Z)$ values)
could be due to the interplay with explicit quark mass effects, which are in principle implementable
in our framework. The precise extrapolation to $\alpha_S(m_Z)$ is however beyond the present scope and postponed
for a future work.
%

\end{document}